\documentclass[twocolumn,showpacs,nofootinbib,tightenlines,nobibnotes, aps, amssymb,amsmath,prl]{revtex4}
\usepackage{graphicx}% Include figure files
\usepackage{epsfig}% Include figure files

\newcommand{\be}{\begin{eqnarray}}
\newcommand{\ee}{\end{eqnarray}}

\begin{document}
\title{Supersolidity in the triangular lattice spin-1/2 XXZ model: A variational perspective}
\author{Dariush Heidarian}
\affiliation{Department of Physics, University of Toronto, Toronto, Ontario, 
Canada M5S 1A7}  
\author{Arun Paramekanti}
\affiliation{Department of Physics, University of Toronto, Toronto, Ontario, 
Canada M5S 1A7} 
\date{\today}

\begin{abstract}
We study the spin-1/2 XXZ model on the triangular lattice with a nearest neighbor
antiferromagnetic Ising coupling $J_z \!> \!0$ and unfrustrated ($J_\perp<0$) or frustrated ($J_\perp>0$)
kinetic terms in zero magnetic field. Incorporating long-range Jastrow correlations over a mean field spin state,
we obtain the variational phase
diagram of this model on large lattices for arbitrary $J_z$ and either sign of $J_\perp$. For $J_\perp\!<\!0$,  we find a
$\sqrt{3}\!\times\!\!\sqrt{3}$
supersolid for $J_z/|J_\perp| \!\gtrsim \!4.7$, in excellent agreement with quantum Monte Carlo data. 
For
$J_\perp \!>\! 0$,
a distinct $\sqrt{3}\!\times\!\!\sqrt{3}$ supersolid is found to emerge for $J_z/J_\perp \!\geq\! 1$. Both
supersolids exhibit a spontaneous density deviation from half-filling.
At $J_z/J_\perp\!\!=\!\!\infty$, the crystalline order parameters of these two supersolids are
nearly identical,
consistent with exact results.
\end{abstract}
\pacs{75.10.Jm 05.30.Jp 71.27.+a}
\vskip2pc

\maketitle

{\it{Introduction. ---}}
Understanding how quantum effects
select a unique ground state from an exponentially large number of
classically degenerate configurations is a problem that is common to
geometrically frustrated magnets, Mott insulators, and quantum Hall systems.
The simplest  example of a frustrated classical 
magnet is the two-dimensional (2D)
classical Ising antiferromagnet on the triangular lattice, which has an extensive ground state
entropy, with an entropy per spin $\approx 0.323$, and critical $\sqrt{3}\times\!\sqrt{3}$
spin correlations \cite{classical}.
The application of an infinitesimal transverse magnetic field 
is known to lift this ground state degeneracy and favor a long-range $\sqrt{3}\times\!\sqrt{3}$
ordered state \cite{blankschtein} where the sublattice
magnetizations on the three sublattices take on the form $(0,m_z,-m_z)$. Another quantum
variant is the
triangular lattice spin-1/2 XXZ model, described by the Hamiltonian
\begin{equation}
H = \sum_ {\langle ij\rangle}[J_zS_i^zS_j^z+J_\perp(S_i^x S_j^x +
S_i^y S_j^y)],
%\label{eq:H_e}
\end{equation}
where $J_z>0$ and $\langle ij\rangle$ refer to nearest neighbour links of the
triangular lattice; interest in this model stems from an early suggestion \cite{fazekas} 
that this model might support, for $J_\perp>0$, a quantum spin
liquid ground state for $J_z/J_\perp \gg 1$. (Henceforth, we will set $|J_\perp|=1$.)
This model may also be viewed as a hard-core boson model with a
nearest neighbor boson repulsion and a boson hopping amplitude 
which is unfrustrated ($J_\perp < 0$) or frustrated ($J_\perp > 0)$. Interest
in such models has been revived following reports \cite{kimchan}
of supersolidity in $^4$He which remain to be understood.

In this paper, using a variational wave 
function which incorporates long range Jastrow correlations, together with
recently developed Monte Carlo optimization methods \cite{vmc}, we obtain the
variational phase diagram of this model on large lattice sizes for either sign of $J_\perp$.
The main
significance of our work, apart from the fact that it represents substantial progress in
the study of the XXZ model, is that it illustrates the power of the variational wave function
approach in studying strongly correlated phases and quantum phase transitions.

We begin by summarizing our key results
in the context of recent important work on this model. 

\underline {For $J_\perp \!<\! 0$:} (i) We incorporate
Jastrow correlations above a uniform superfluid wave function
and find, using the Binder cumulant, 
a phase transition from a correlated superfluid state into a supersolid with long range 
$\sqrt{3}\times\!\sqrt{3}$ crystal order for $J_z \!\gtrsim\! 4.7$;  this supersolid order persists 
even for $J_z \!\to\! \infty$. 
This is in good agreement with quantum Monte Carlo (QMC) simulation results 
\cite{dariush_kedar,melko,troyer}
which found a 
superfluid-supersolid transition at $J_z \approx 4.5$. (ii) We find that the Jastrow factors exhibit
a long range $1/r$ tail in the superfluid as well as the supersolid which is consistent
with a linearly dispersing sound mode. (iii) The crystal order in the supersolid
emerges as a {\it spontaneously}
broken symmetry in our wave function as a consequence of Jastrow correlations, 
but 
%we show that 
the Jastrow factors do not
completely fix the crystal pattern. Incorporating a variational parameter which directly
couples to the
difference between the $(2m_z,-m_z,-m_z)$ and $(0,m_z,-m_z)$ candidate crystal orders, we find
that
the former has a slightly lower energy, consistent with QMC results 
\cite{dariush_kedar,troyer,boninsegni}
as well as a recent variational study \cite{moessner} of the quantum dimer model at $J_z\!=\!\infty$.
(iv) We find that a supersolid wave function which permits for a small density deviation away
from half-filling has lower energy.

\underline{For $J_\perp > 0$:} (i) We incorporate Jastrow correlations on top of a state with
coplanar 3-sublattice magnetic order. Using this wave function, we find that there is
$120^\circ$-xy order 
for $J_z \!<\! 1$ whereas there is supersolid
order for $J_z \gtrsim 1$, with $(2m_z,-m_z,-m_z)$ crystal order,
in agreement with a recent suggestion based on analyzing  the $J_z\!=\!\infty$ limit
\cite{pollmann} and density matrix renormalization group (DMRG)
results on small system sizes \cite{DMRG}. (ii) The formation of the supersolid state is accompanied 
by a spontaneous
density deviation away from half-filling. Equivalently, the XXZ model in zero field develops a 
spontaneous $S^z$ magnetization; this magnetization grows with increasing $J_z$. (iii) The
$120^\circ$-xy state as well as the supersolid exhibit $1/r$ decay of Jastrow factors consistent with
a linearly dispersing excitation mode.
(iv) For $J_z\!=\!\infty$, recent work has shown,
through the construction of a
unitary transformation \cite{pollmann}
as well as by a high temperature expansion \cite{DMRG},
that
the diagonal spin correlations are identical for either sign of $J_\perp$
in this limit, which enables one
to conclude that the frustrated XXZ model must also exhibit $\sqrt{3}\!\times\!\sqrt{3}$ crystal order, 
and is likely to be a supersolid at $J_z\!=\!\infty$ \cite{pollmann,DMRG}.
At $J_z\!=\!\infty$, the crystal orders in our `frustrated supersolid' and the
`unfrustrated supersolid' are nearly identical and
close to QMC results for $J_\perp\!<\!0$, in agreement with this result. (v) In contrast to the
result of Ref.~\cite{pollmann}, we find that the off-diagonal order parameter for $J_\perp \!>\! 0$
is much smaller than that for $J_\perp\! < \! 0$.

% (v)  For $J_z<1$, the variational
%ground state has planar $120^\circ$ order and its energy is in good agreement with previous 
%exact diagonalization results.

{\it Variational Wave function. ---}
We begin with a variational wave function of the form
\be
|\psi_{\rm V}\rangle= 
{\cal J}_{\rm \theta}
\exp\bigl(\sum_{i, {\bf r}} {\cal J}({\bf r}) S_i^z S_{i+{\bf r}}^z\bigr)
\mathcal{P}_{\rm G}
|\psi_{\rm MF}\rangle,
\ee 
where
$|\psi_{\rm MF}\rangle$
is a variational mean-field wave function for spin-$\uparrow$ and spin-$\downarrow$ fermions (see below), 
$\mathcal{P}_{\rm G}$ is the Gutzwiller projection operator which restricts the Hilbert space to one
fermion per site thus taking us from a fermionic state to a spin wave function,
${\cal J}({\bf r})$ are variational Jastrow parameters which build correlations between spins separated by ${\bf r}$ 
while respecting lattice symmetries, and ${\cal J}_\theta$ is a global variational parameter, elaborated upon
later, which fully fixes the crystalline order.

We choose
$|\psi_{\rm MF}\rangle$ to be the ground state of a mean field Hamiltonian
$
H_{\rm MF}= - \sum_{\langle ij\rangle}t^{\rm var}_{ij}c_{i\sigma}^\dagger c_{j\sigma} - \sum_i{\vec h^{\rm var}_i}\cdot {\vec S_i},
\label{eq:MF}
$
where $\vec S_i = \frac{1}{2} c_{i\alpha}^\dagger \vec\sigma_{\alpha\beta} c_{j\beta}$.
The first term in $H_{\rm MF}$ represents nearest neighbor hopping of fermions, with amplitudes
$t^{\rm var}_{ij}$, while
the second term represents a site-dependent magnetic field
whose direction and magnitude can be chosen to yield various magnetically ordered states.
Note that the wave function has the freedom to
decribe a magnetically disordered spin liquid state if $h^{\rm var}_i \!=\! 0$.

For the unfrustrated model ($J_\perp\!=\!-1$) we set $t^{\rm var}_{ij}=0$, and select two different
configurations of $\vec h^{\rm var}_i$. One choice is to set  $\vec h_i^{\rm var}
= \hat{x}$, which leads to a mean field ferromagnet
with the moment polarized along the $S^x$
direction. This mean field state has no variational parameters; we refer to the spin wave function, 
after Gutzwiller projecting this state and inclusion of correlation factors, as 
$|\Psi_V^{\rm FM-x}\rangle$. The second choice, which we motivate below, promotes 3-sublattice
magnetic order, with $h_{i,z}^{\rm var}=(1,-1/2,-1/2)$ on the 3 sublattices,
$h^{\rm var}_{i,x}=h_x$, and $h^{\rm var}_{i,y}=0$,
where $h_x$ is a variational parameter.
We refer to the correlated spin state obtained from this mean field
state as $|\Psi_V^{\rm FM-xz}\rangle$.

For the frustrated model, we choose the
hopping amplitudes such that $|t^{\rm var}_{ij}|=1$ but with signs picked such that upward (downward) 
pointing triangular plaquettes enclose
flux $\pi$ (zero). Upon Gutzwiller projection, this state describes a spin liquid which preserves all
lattice symmetries. It is equivalent
to a pairing wave function which is known, from earlier studies \cite{vmc}
to be a good
starting point to understand the triangular Heisenberg antiferromagnet. Motivated by the
classical ground state, we pick $\vec h^{\rm var}_i$ to describe
3-sublattice coplanar magnetic order. The spin wave function obtained from choosing
$h^{\rm var}_{i x}=(h,-h/2,-h/2)$, $h^{\rm var}_{i y}=(0, h \sqrt{3}/2, - h \sqrt{3}/2)$, and
$h^{\rm var}_{iz}=0$, with $h$ being a mean field variational parameter,
will be called $|\Psi_V^{\rm AF-xy}\rangle$.
The spin wave function obtained from choosing
$h^{\rm var}_{i x}=(0,h_x,-h_x)$, $h^{\rm var}_{i z}=(h_{z}, - h_z/2,-h_z/2)$, and
$h^{\rm var}_{iy}=0$, with mean field variational parameters $h_x,h_z$, will be called
$|\Psi_V^{\rm AF-xz}\rangle$.

In order to obtain the best trial wave function, we optimize the variational parameters 
using Monte Carlo sampling of the spin configurations combined with recently developed
techniques \cite{vmc}.
We begin by discussing our results for 
the unfrustrated model, for which we present comparisons with
existing QMC results. We follow it up with results on the frustrated model
which is less well studied since the sign problem does not permit controlled QMC simulations.
Finally, we consider  $J_z=\infty$, and demonstrate consistency with exact results in this limit.

\begin{figure}[t]
\includegraphics[width=.45\hsize]{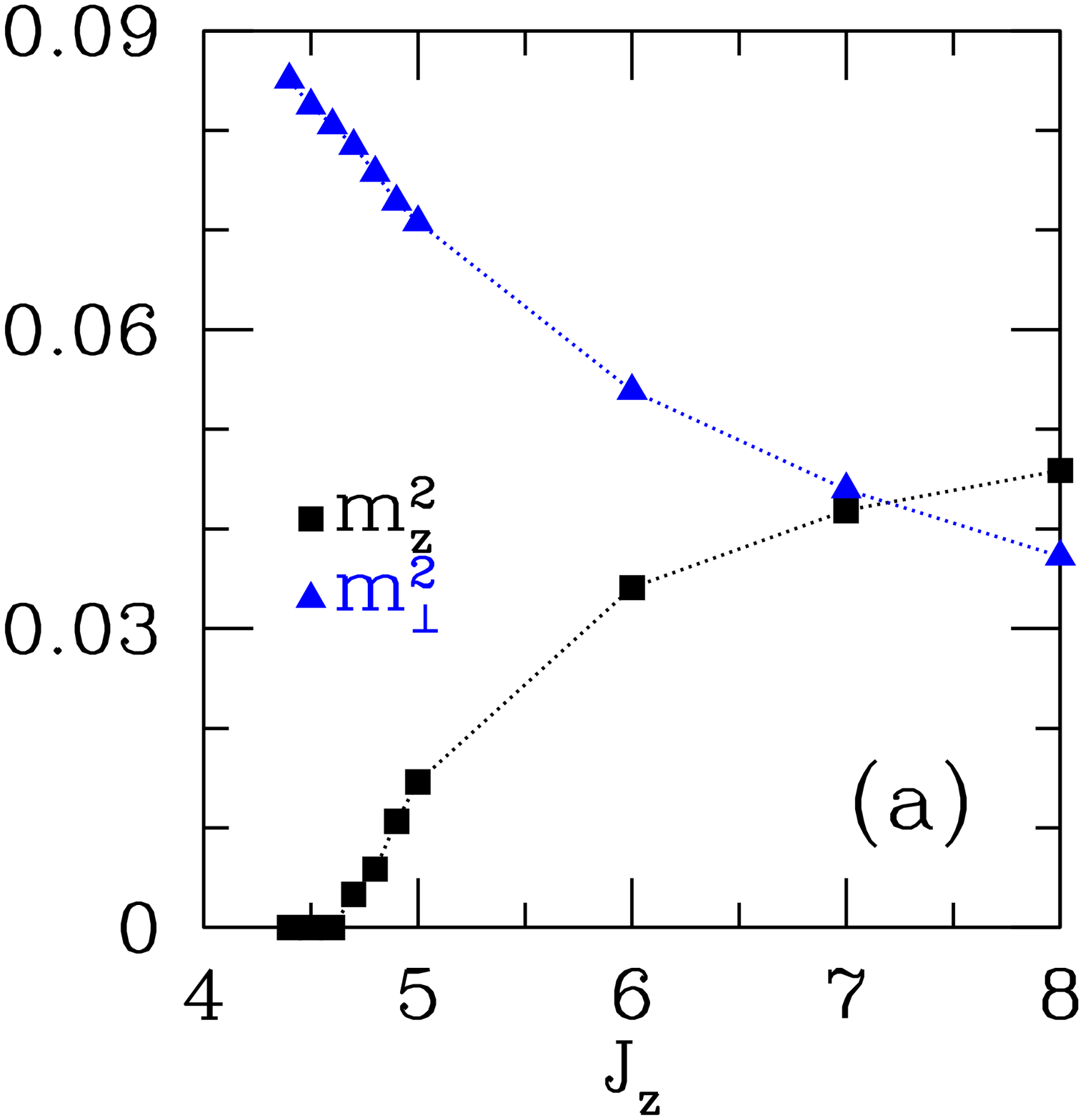}
\includegraphics[width=.45\hsize]{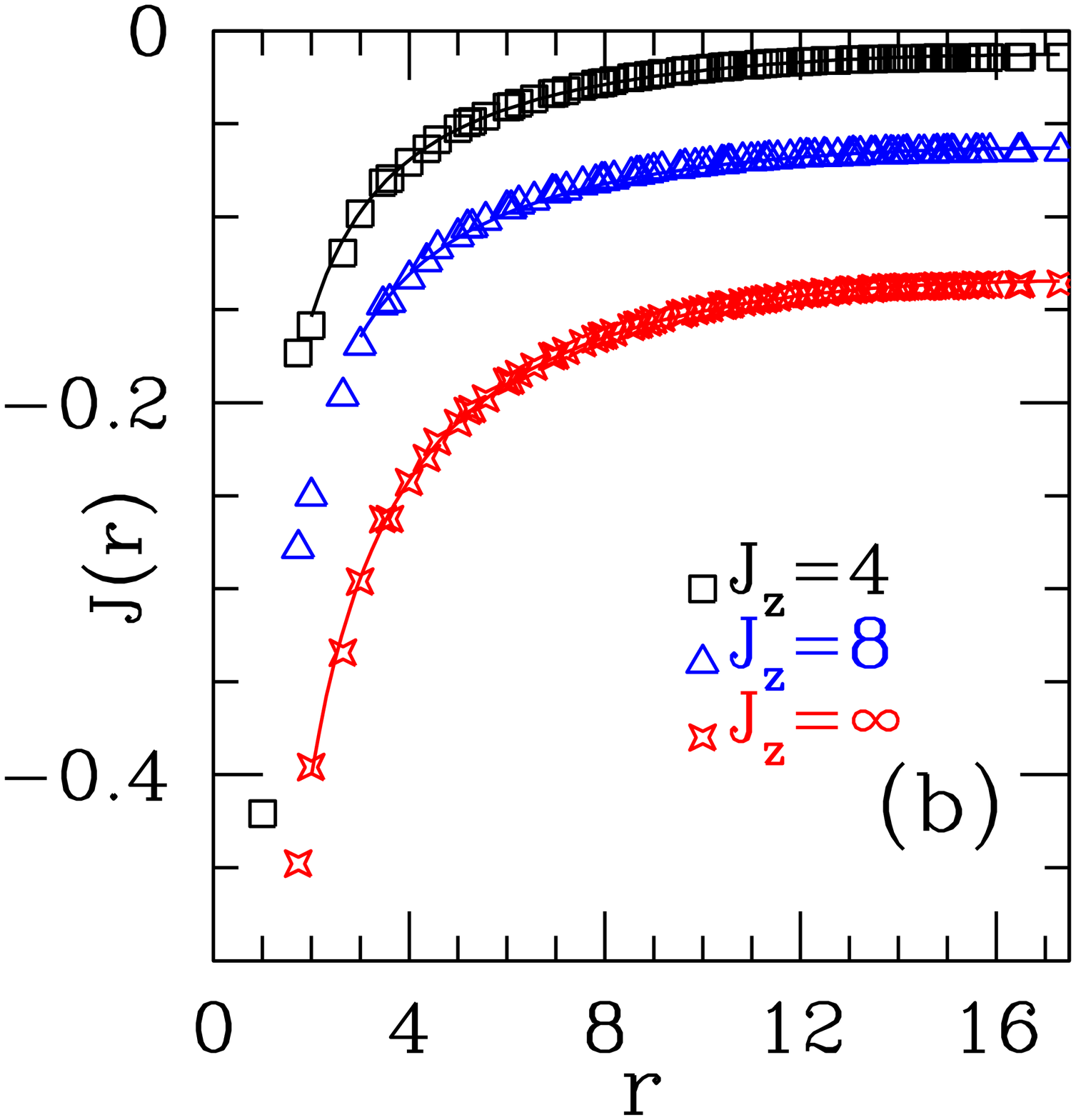}
\caption{(Color online) (a) Order parameters, extrapolated to the thermodynamic limit,
for $J_\perp\!=\!-1$. (b) Jastrow parameters showing $1/r$ behavior at long distance (curves for
$J_z=4,\infty$ have been 
offset downward for clarity).}
\label{m2FM_jastrow}
\end{figure}

{\it $J_\perp < 0$, Unfrustrated Case. ---} For $J_z \ll 1$, the ground state of the
unfrustrated model is a quantum ferromagnet or,
equivalently, a superfluid state of hard core bosons. 
With increasing $J_z$, the ground state of the superfluid becomes increasingly correlated, and 
QMC simulations 
\cite{dariush_kedar,melko,troyer,boninsegni}
show that a supersolid ground state, with coexisting superfluid
and charge order, develops for $J_z \!\gtrsim\! 4.5$. If we set ${\cal J}_\theta=1$ and optimize 
$|\Psi_V^{\rm FM-x}\rangle$,
we find that the ferromagnetic order parameter $m^2_\perp = |\langle S^x \rangle|^2$ 
is nonzero for all values
of $J_z$, as seen from Fig.~\ref{m2FM_jastrow}(a), consistent with superfluidity persisting for arbitrarily 
large interactions.
For all
interaction strengths, we also find, as seen from Fig.~\ref{m2FM_jastrow}(b), 
that  the optimal Jastrow parameters ${\cal J}({\bf r})$ decay 
as $1/r$ with distance. Such a power law decay implies a density structure
factor scaling as $\sim |{\bf q}|$ at small momenta which is consistent with a linearly dispersing
Goldstone mode above a superfluid ground state \cite{trivedi}.
It is a nontrivial test of the unbiased variational optimization 
that we recover this behavior by optimizing over a very large set of variational Jastrow parameters.

\begin{figure}[t]
\includegraphics[width=.4\hsize]{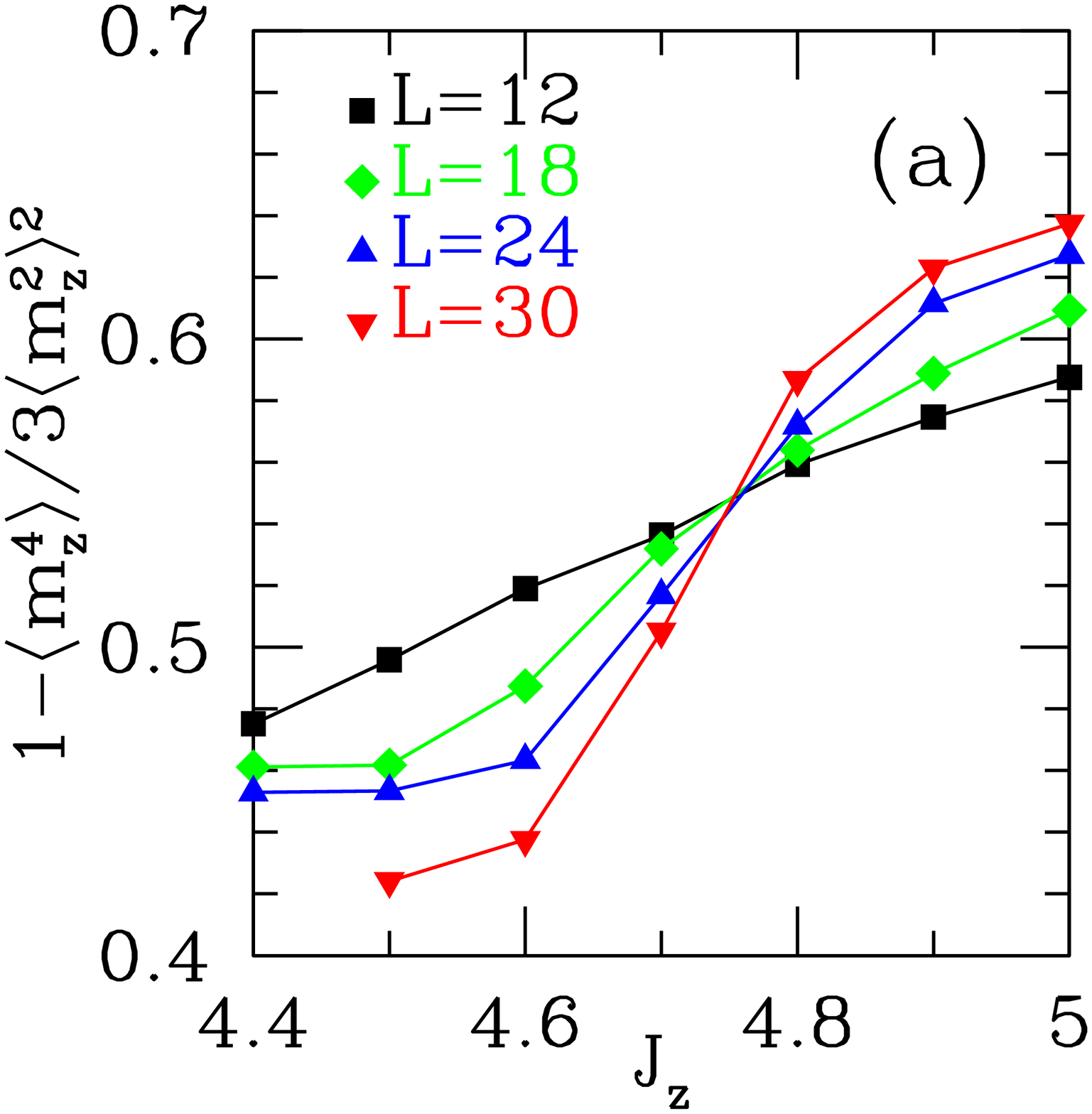}
\includegraphics[width=.4\hsize]{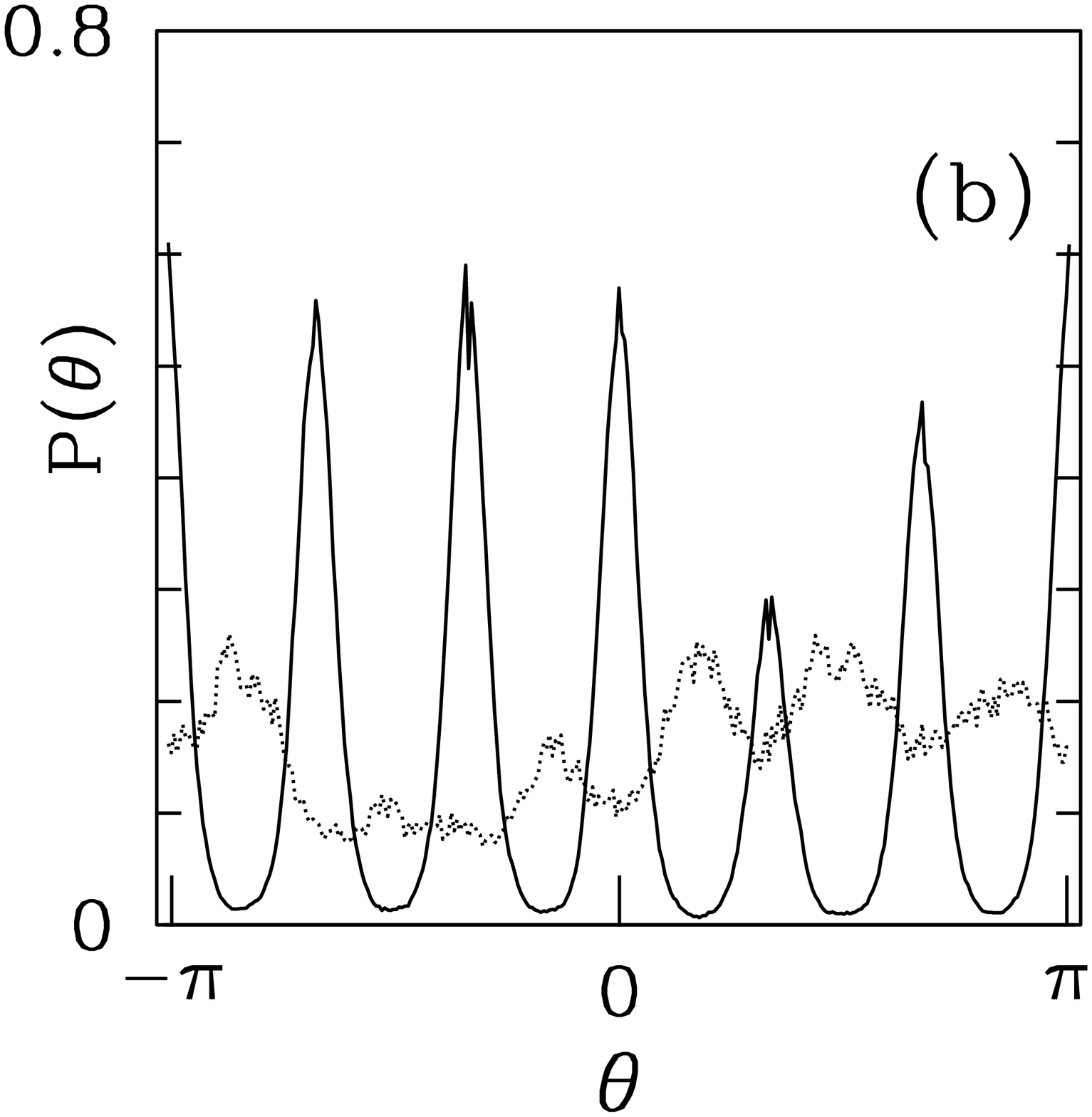}
\caption{(Color online) (a) Binder cumulant crossing for $m_z$ displaying a superfluid-supersolid transition at 
$J_z^{\rm (crit)} \! \approx \! 4.7$.
(b) Probability distribution of the phase of order 
parameter for $h_\theta\!=\!0$ (dashed line) and at the optimal $h_\theta\!>\!0$ (solid line).}
\label{binder_phase}
\end{figure}

In order to probe for supersolidity, we compute the density structure factor (equivalently
$S^z$ correlations). Although $|\Psi_V^{\rm FM-x}\rangle$ is a translationally invariant
state, we find that the density structure factor diverges at momenta $\pm (4\pi/3,0)$ for large $J_z$
which shows that there is $\sqrt{3}\times\sqrt{3}$ crystalline ordering, that coexists with the
superfluidity, for strong interactions. 
It is remarkable that the crystalline order emerges as a {\it spontaneously broken symmetry} induced by
Jastrow correlations in our wave function.
Using the crossing of the Binder cumulant on different system 
sizes as shown in Fig.~\ref{binder_phase}(a), as well as from a calculation of the order parameters extrapolated to
infinite system sizes shown in Fig.~\ref{m2FM_jastrow}(a), we locate the superfluid to supersolid transition point at
$J^{\rm (crit)}_z \!\approx\! 4.7$, which is in good agreement with QMC simulation results.

While the crystal order is spontaneously generated in this wave function, a complete characterization of the 
crystalline order requires knowledge of the
phase of the order parameter 
%\be
$ \Phi=S^z_A+S^z_B {\rm e}^{i 2\pi/3}+S^z_C
{\rm e}^{i 4\pi/3}
\equiv |\Phi| {\rm e}^{i\theta},
\label{op}
$
%\ee 
where $S^z_{A,B,C}$ refer to the total density on the $A,B,C$ sublattices.
Specifically the two candidate  $\sqrt{3}\times\sqrt{3}$ crystal orders for this model have
$\theta= 2 n \pi/6$ with $n=0,\cdots,5$, which corresponds to having a density order
$(2 m_z,-m_z,-m_z)$ on the three sublattices, and $\theta= (2 n + 1) \pi/6$ with $n=0,\cdots,5$,
which implies $(0,m_z,-m_z)$ order. Since the Jastrow factors are insensitive to the phase of $\Phi$, they
cannot select between these candidates. In order to find the
optimal supersolid, we therefore include a factor ${\cal J}_\theta=\exp( h_\theta \cos 6\theta)$ in the
wave function, with $h_\theta$ being a variational parameter. Optimizing this variational parameter,
we find that $h_\theta > 0$ has the lowest energy indicating that the best supersolid state has
$(2 m_z,-m_z,-m_z)$ crystal order.
As shown in Fig.~\ref{binder_phase}(b) for $J_z=\infty$, 
the probability distribution $P(\theta)$, of the phase $\theta$,
shows sharp peaks at the expected angles at the 
optimal $h_\theta>0$ \cite{unbiased},
while the state with $h_\theta\!=\!0$ is relatively featureless
(the small peaks at $\theta= (2 n + 1) \pi/6$ appear to be a finite size artifact).
Our results are consistent with earlier QMC simulation results \cite{dariush_kedar,troyer,boninsegni}
and recent
variational calculations \cite{moessner} in the quantum dimer model limit at $J_z=\infty$.
Motivated by these results, we study the properties of $|\Psi_V^{\rm FM-xz}\rangle$
which imposes the above supersolid order, and find that the optimized wave function of
this type exhibits a small density deviation away from half-filling. Equivalently, the
XXZ model has a spontaneous $S^z$ magnetization in the supersolid.

\begin{table}
\caption{\label{table}
Exact diagonalization energy \cite{leung} versus variational energy of $|\Psi_V^{\rm AF-xy}\rangle$ 
for a $6\!\times\!6$ system with $J_z \leq 1$.
}
\begin{ruledtabular}
\begin{tabular}{l|llll}
%J_z$ & VMC & ED \\
$J_z$ & $0.0$ & $0.5$ & $0.9$ & $1.0$\\
\hline
VMC & -0.40950(7) & -0.46992(5) & -0.53075(5) & -0.54816(9)\\ 
ED & -0.410957 & -0.475366 & -0.541988 & -0.560373\\
%$0.0$ & -0.40948 & -0.410957 \\
%$0.5$ & -0.46992 & -0.475366 \\
%$0.75$ &-0.50650 & -0.515612 \\
%$0.9$ & -0.53075 & -0.541988 \\
%$1.0$ & -0.54800 & -0.560373 \\
\end{tabular}
\end{ruledtabular}
\label{table_energy}
\end{table}

\begin{figure}[t]
\includegraphics[width=.4\hsize]{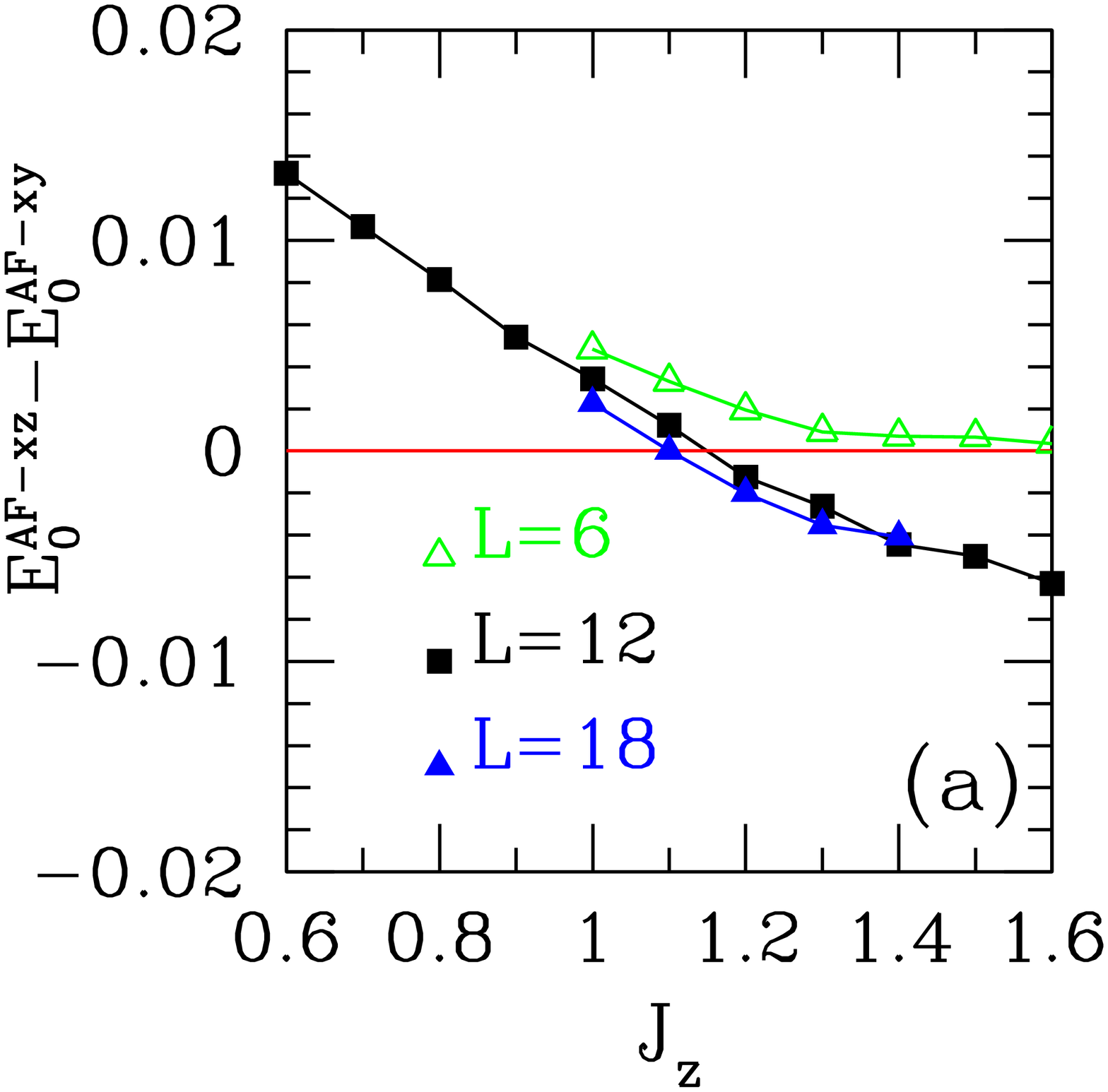}
\includegraphics[width=.4\hsize]{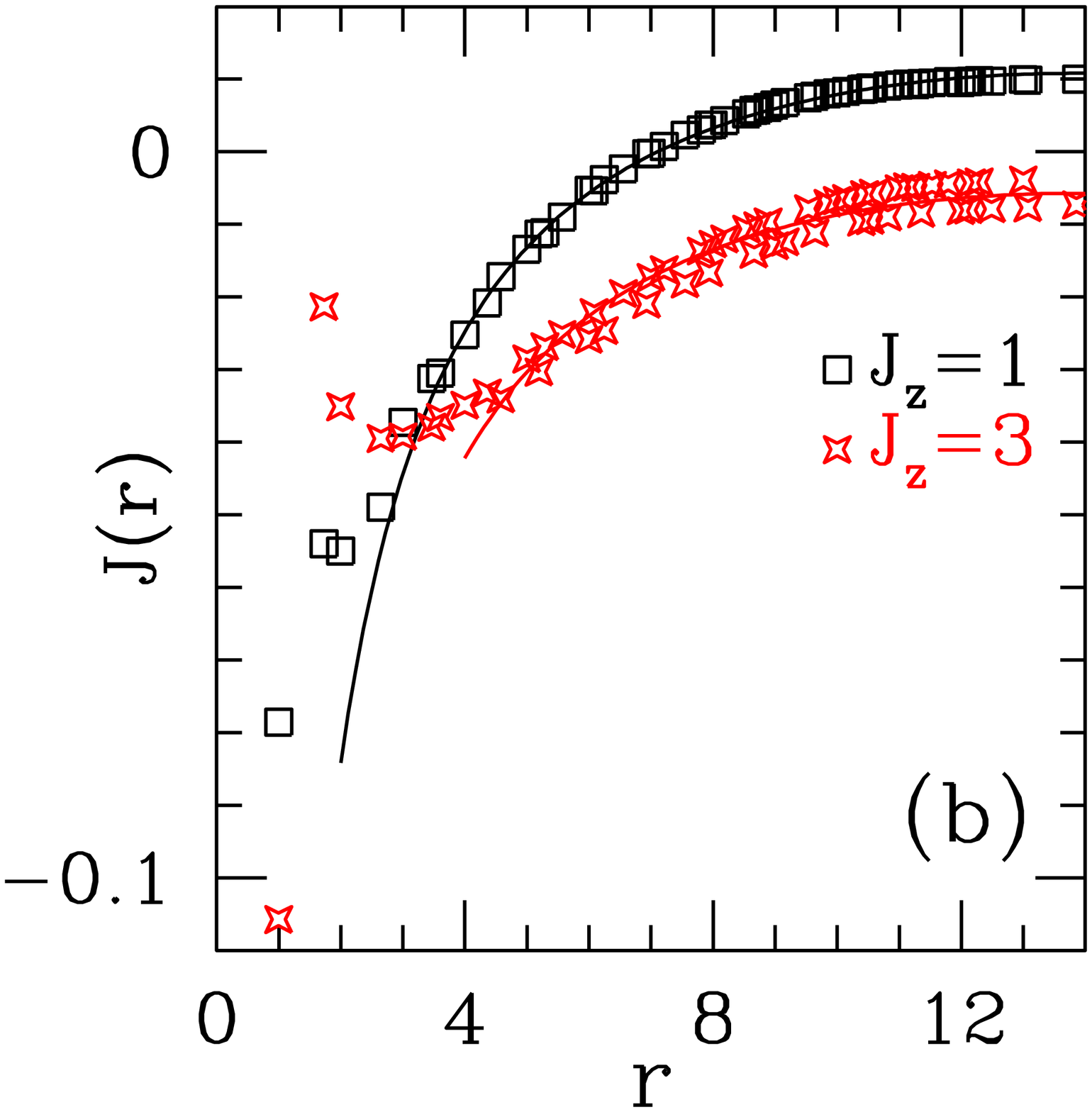}
\caption{(Color online)
(a) Energy difference $\Delta E = E_0^{\rm AF-xz}-E_0^{\rm AF-xy}$
showing
that $|\psi_{V}^{\rm AF-xz}\rangle$ has lower energy for $J_z \!\gtrsim\! 1$.
(b) Jastrow parameters versus distance for $J_z\!=\!1$ ($|\psi_{V}^{\rm AF-xy}\rangle$) and
$J_z\!=\!3$ ($|\psi_{V}^{\rm AF-xz}\rangle$).
Solid line depicts $1/r$ behavior. Curve for $J_z=1$ has been shifted upward for clarity.}
\label{jastrow_energy}
\end{figure}

{\it $J_\perp > 0$, Frustrated Case. ---}
For the frustrated model, the sign problem prevents controlled QMC simulations on large lattices. There
have been some classical and spin wave
analyses for large spin \cite{largeS}, and exact diagonalization (ED) studies on small systems \cite{leung}
which do not address the issue of the thermodynamic limit for spin-$1/2$.
The classical ground state with $J_z \!<\! 1$ has $120^\circ$ magnetic ordering in the ${\rm xy}$-plane.
We therefore begin with the wavefunction $|\Psi_V^{\rm AF-xy}\rangle$ as a variational candidate. 
With increasing $J_z$, the system becomes more strongly correlated, developing strong short
range $S^z$ correlations. Table I shows that the variational energy compares well with
ED results  \cite{leung} on a $6\!\times \!6$ lattice for $J_z \! \leq \! 1$.

From the density
structure factor and probability distributions of the phase of $\Phi$, defined earlier,
we find that this wave function
also displays a spontaneously ordered $\sqrt{3}\times\sqrt{3}$ supersolid, with staggered charge and 
superfluid orders, for $J_z\!
\gtrsim\! 1.2$. The
charge order is of the type $(2 m_z, -m_z, -m_z)$ on the three sublattices, identical to what we found in the
unfrustrated model, and $m_z$ grows continuously beyond $J_z \!\approx\! 1.2$, whereas
the superfluid order is of the type $(0,m_\perp,-m_\perp)$ on the 3 sublattices.
Motivated by the proximity
of this transition to the Heisenberg model,
we compare the two states, $|\Psi_V^{\rm AF-xy}\rangle$
and $|\Psi_V^{\rm AF-xz}\rangle$, to see if the plane of the coplanar order rotates at this SU(2) symmetric
point.
As shown in Fig.~\ref{jastrow_energy}(a), the energy of $|\Psi_V^{\rm AF-xz}\rangle$ is indeed lower for
$J_z \!\gtrsim \!1$.
We find that the Jastrow factors in $|\Psi_V^{\rm AF-xy}\rangle$ as well as $|\Psi_V^{\rm AF-xz}\rangle$ 
exhibit a long range $1/r$ tail consistent
with 
%zero point fluctuations arising from 
a linearly dispersing mode
(see Fig.~\ref{jastrow_energy}(b)).

Using the two different wave functions on the two sides of the Heisenberg point, we 
conclude that $m_z$ and $m_\perp$ are discontinuous at $J_z\!=\!1$, and $|\Psi_V^{\rm AF-xz}\rangle$
describes a supersolid state
% in this model for $J_z \!>\! 1$ 
as seen from Fig.~4(a).
For a $6\!\times\!6$ system, our results are in excellent agreement with DMRG calculations
\cite{DMRG} (see Fig.~4(b)).
The crystal order grows monotonically for $J_z \!>\!1$, while $m_\perp$ gets
strongly suppressed. We find that $|\Psi_V^{\rm AF-xz}\rangle$ exhibits a small spontaneous
density deviation
from half-filling, as shown in Fig.~4(c), which grows with $J_z$ to a
maximum value $\approx 0.05\%$.

\begin{figure}[t]
\includegraphics[width=.4\hsize]{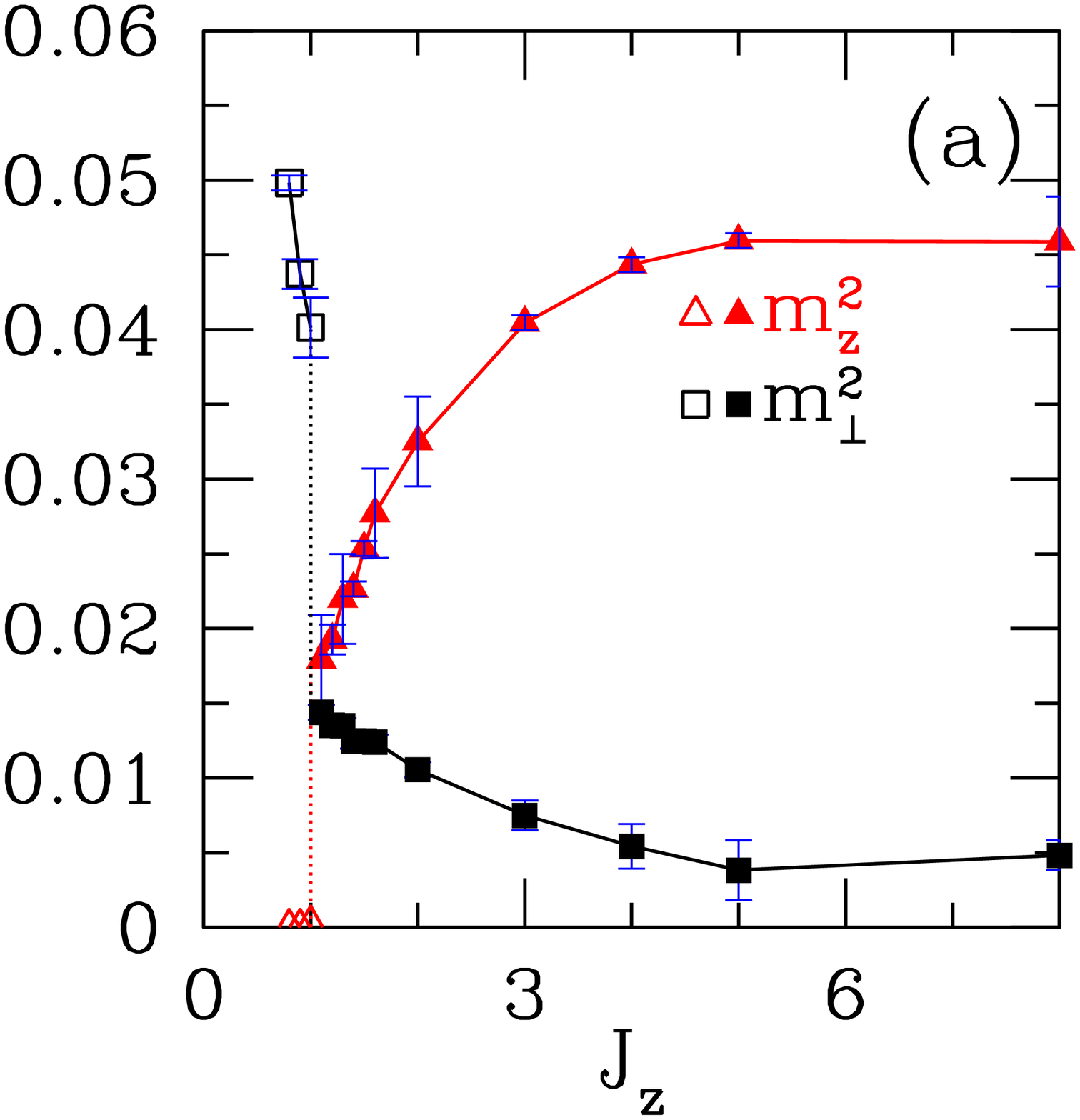}
\includegraphics[width=.4\hsize]{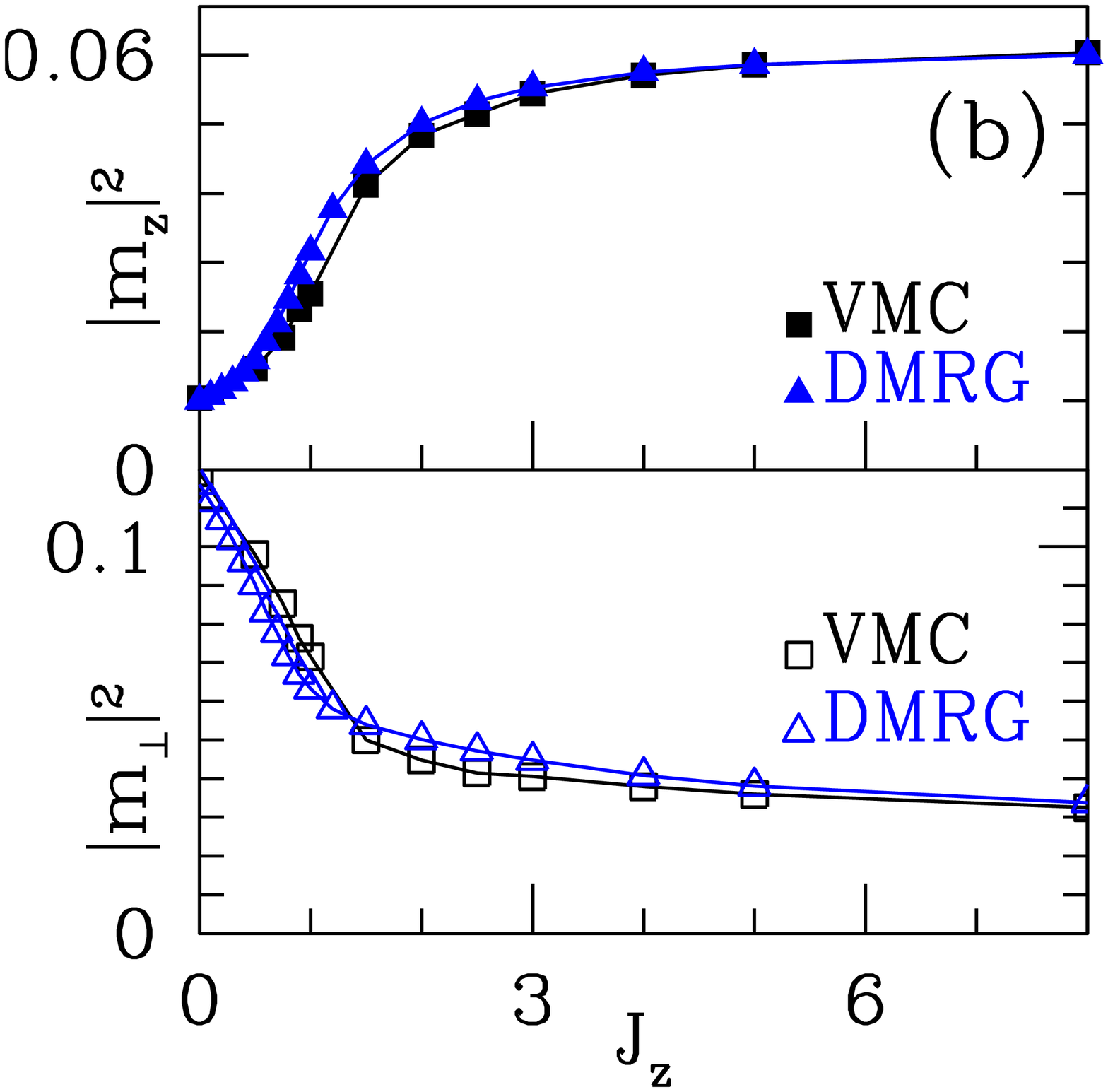}
\includegraphics[width=.4\hsize]{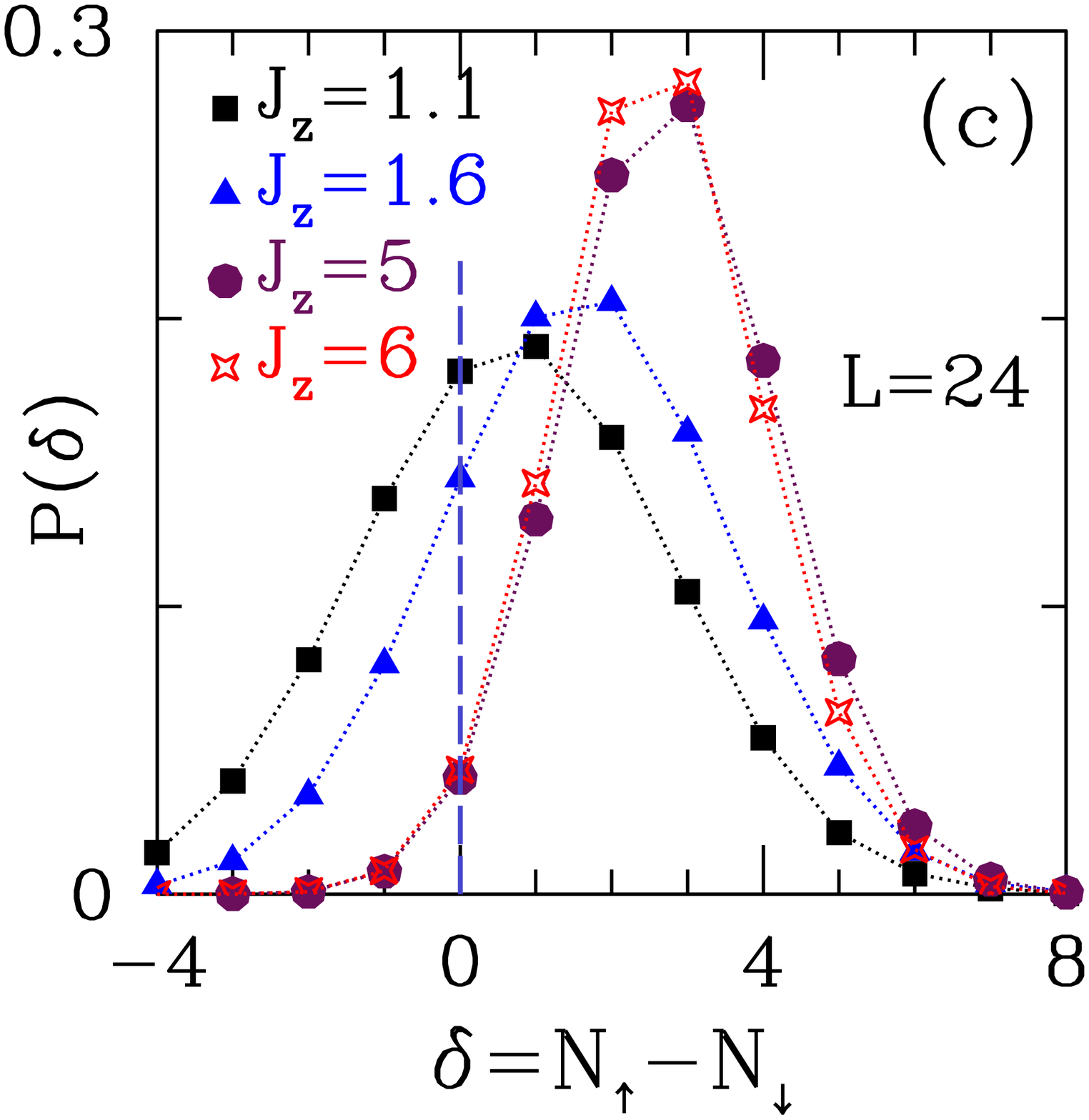}
\includegraphics[width=.4\hsize]{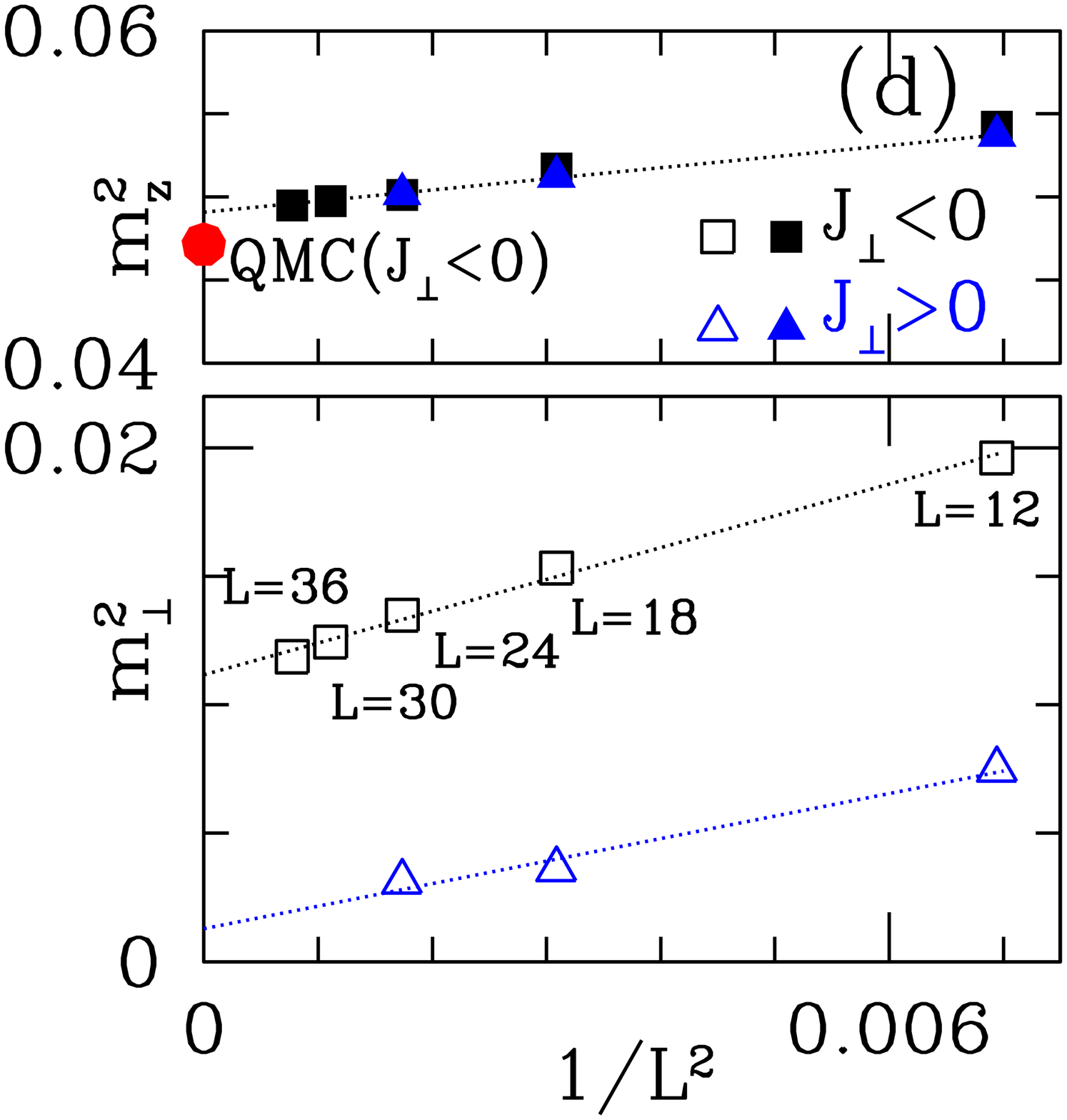}
\caption{(Color online)
(a) Diagonal and off-diagonal order parameters extrapolated to the thermodynamic limit showing the order
parameter discontinuity at $J_z \!\simeq\! 1$ and the coexistence of both orders for $J_z \!>\!1$. 
(b) Order parameters
for a $6\times 6$ system compared with DMRG results from Ref.\cite{DMRG}.
(c) Density distribution in $|\Psi_V^{\rm AF-xz}\rangle$ showing nonzero magnetization with the onset of
supersolid order. (d) Diagonal and off-diagonal order parameters in the limit $J_z\!=\!\infty$ showing
good agreement of the extrapolated $m_z^2$ for $J_\perp>0$ with QMC data on the unfrustrated model.}
\label{Jzinf}
\end{figure}

{\it $J_z\!=\!\infty$, the Quantum Dimer Limit. ---}
At $J_z\!=\!\infty$, the Hilbert space of the XXZ model gets constrained to the
minimally frustrated classical configurations of the triangular lattice
Ising antiferromagnet \cite{classical}. 
%Remarkably, there is a nonlocal unitary transformation \cite{pollmann}, which allows one to change the
%sign of $J_\perp$, and shows that the spectrum of the XXZ model
%must be identical for $J_\perp\!=\!\pm 1$, in this limit. This mapping also leaves the $S^z$ correlations
%invariant, so that the frustrated and unfrustrated models must have identical crystalline order in this limit.
Remarkably, although the `frustrated supersolid' wave function $|\Psi_V^{\rm AF-xz}\rangle$ describes a 
very different state from the `unfrustrated supersolid' $|\Psi_V^{\rm FM-x}\rangle$,  we find, upon projecting
to the minimally frustrated Ising subspace,
that the crystal order parameters of both states are nearly equal, and are in good
agreement with the extrapolated
QMC result for the unfrustrated case, as shown in Fig.~\ref{Jzinf}(d). We also find that the energies
of these two completely different states are fairly close, within $6\%$ of each other. This is consistent 
with recently uncovered
exact results \cite{pollmann,DMRG} at $J_z\!=\!\infty$ which show that the crystal order parameters and
ground state energies should be equal for either sign of $J_\perp$. As seen from Fig.~4(d), $m_\perp$ 
extrapolates to a small value for either sign of $J_\perp$, with $m_\perp(J_\perp\!=\!1) \!\ll\! m_\perp (J_\perp\!=\!-1)$.

{\it Summary. ---} In summary, we have used a variational approach to obtain the phase diagram of
the frustrated and unfrustrated spin-1/2 XXZ models on the triangular lattice for large lattice sizes and 
arbitrary $J_z>0$;
these regimes are not accessible via other techniques. The phase diagram of the unfrustrated
model is in good agreement with QMC data. The frustrated model
exhibits a wide regime of supersolidity with no evidence of a spin liquid. Our results
for $J_\perp < 0$ could be tested using polarized dipolar bosonic molecules in a triangular lattice by
tuning the ratio $J_\perp/J_z$. For the frustrated model, proposed schemes to induce
fictitious gauge fields \cite{gauge}
need to be adapted to produce a frustrated molecule hopping.

{\it{Acknowledgements:}} We acknowledge support from
NSERC of Canada, the A. P. Sloan Foundation, the Ontario ERA, and the Connaught
Foundation.


\begin{thebibliography}{999}
\bibitem{unbiased}
Our optimization scheme is sufficiently unbiased that even if we set the initial
$h_\theta < 0$, it changes sign after several optimization steps and leads to 
an optimal $h_\theta>0$.
\end{thebibliography}
\end{document}